\documentclass[prb,aps,twocolumn,showpacs,amsmath,amssymb,floatfix]{revtex4}

\usepackage{graphicx}

\begin{document}
%
%%%%%%%%%%%%%%%%%%%%%%%%%%%%%%%%%%%%%%%%%%

\newcommand{\rmi}{\mathrm{i}}
\newcommand{\etal}{{\it et al}}
%\newcommand{\eqref}[1]{({\ref{#1}})}

% My commonly quoted journals
%
\newcommand{\IJMPB}{Int. J. Mod. Phys. B }
\newcommand{\PhC}{Physica C }
\newcommand{\PhB}{Physica B }
\newcommand{\JS}{J. Supercond. }
\newcommand{\IEEEmw}{IEEE Trans. Microwave Theory Tech. }
\newcommand{\IEEEas}{IEEE Trans. Appl. Supercond. }
\newcommand{\IEEEim}{IEEE Trans. Instr. Meas. }
\newcommand{\PRB}{Phys. Rev. B }
\newcommand{\PRL}{Phys. Rev. Lett. }
\newcommand{\PR}{Phys. Rev. }
\newcommand{\IJIMW}{Int. J. Infrared Millim. Waves }
\newcommand{\APL}{Appl. Phys. Lett. }
\newcommand{\JAP}{J. Appl. Phys. }
\newcommand{\JPCM}{J. Phys.: Condens. Matter }
\newcommand{\JPCS}{J. Phys. Chem. Solids }
\newcommand{\AdP}{Adv. Phys. }
\newcommand{\Nat}{Nature }
\newcommand{\CM}{cond-mat/ }
\newcommand{\JpnJAP}{Jpn. J. Appl. Phys. }
\newcommand{\PhT}{Phys. Today }
\newcommand{\ZETF}{Zh. Eksperim. i. Teor. Fiz. }
\newcommand{\JETP}{Soviet Phys.--JETP }
\newcommand{\EL}{Europhys. Lett. }
\newcommand{\Sci}{Science }
\newcommand{\EJPB}{Eur. J. Phys. B }
\newcommand{\IJMB}{Int. J. of Mod. Phys. B }
\newcommand{\RPP}{Rep. Prog. Phys. }
\newcommand{\SUST}{Supercond. Sci. Technol. }
\newcommand{\JLTP}{J. Low Temp. Phys. }
\newcommand{\RSI}{Rev. Sci. Instrum. }
\newcommand{\RMP}{Rev. Mod. Phys. }
\newcommand{\LTP}{Low Temp. Phys. }

\title{Reliable determination of vortex parameters from measurements of the microwave complex resistivity.}
\today

\author{N. Pompeo\footnote[4]{E-mail:pompeo@fis.uniroma3.it.}}
\affiliation{Dipartimento di Fisica ``E. Amaldi'' and Unit\`a CNISM, Universit\`a Roma Tre, Via della Vasca Navale 84, I-00146 Roma, Italy}
\author{E. Silva}
\affiliation{Dipartimento di Fisica ``E. Amaldi'' and Unit\`a CNISM, Universit\`a Roma Tre, Via della Vasca Navale 84, I-00146 Roma, Italy}

\begin{abstract}
We discuss and propose a complete data treatment, in close contact to typical microwave experimental data, in order to derive vortex parameters, such as pinning constant and viscous drag coefficient (also referred to as ``vortex viscosity''), in a way as model-independent as possible. We show that many of the accepted models for the complex resistivity can be described by a single, very general analytical expression. Using typical measurements of real and imaginary resistivity as a function of the applied field, we show that, even for single-frequency measurements, it is always possible to obtain (a) estimates of viscous drag coefficient and pinning constant with well-defined upper and lower bounds and (b) quantitative information about thermal creep. It turns out that neglecting thermal creep, in particular and counterintuitively at low temperatures, might result in a severe overestimation of the viscous drag coefficient. We also discuss the impact of thermal creep on the determination of the pinning constant. The present results might lead to a reconsideration of several estimates of the vortex parameters.
\end{abstract}
\pacs{74.25.Nf, 74.25.Qt, 74.72.Jt, 74.72.Bk}
%\keywords{surface impedance, microwave, vortex dynamics, Tl2212, Y123}

\maketitle

\section{Introduction}
\label{intro}
The electromagnetic response of High-$T_c$ superconductors (HTCS) in the mixed state is of great interest for fundamental physics as well as for technological applications (e.g. pickup coils for magnetic resonance imaging,\cite{mri} filters\cite{filters}). In particular, controlling vortex pinning is essential to reduce power dissipation and signal noise in devices, whereas intrinsic macroscopic quantities such as the viscous drag coefficient are intimately related to the electronic states of vortex cores, and thus yield a great deal of information about the latter.\cite{kopninRPP,maedaJPCM} It is then desirable to obtain accurate and reliable determinations of the vortex parameters from experimental data. As a matter of fact, however, in HTCS the estimates of the vortex parameters span orders of magnitude, even when similar compounds are measured.\cite{GolosovskySUST96}
The HTCS vortex dynamics is notoriously very complex.\cite{Bishop,blatterone,fisher2huse} Thus, the determination of the vortex parameters from the experimental data is not straightforward (to say the least).
Among the others, vortex-vortex interactions, which give rise to non-local response,\cite{OngNonLocal} and disorder \cite{blatterone} play a fundamental role. As a consequence, the response of the vortex system can attain very complex frequency dependencies even in the linear regime of small currents.\cite{WuCorbino,Belk,P&P,MSexp,OngNonLocal,Wuc66} In those cases, it is rather difficult to reliably determine (or even define) the vortex parameters.

A substantial simplification comes by increasing the stimulus frequency to sufficiently high values, e.g. to the microwave range. In this case the amplitude of the vortex oscillations becomes so small\cite{tomasch} that the system can be treated in the local, single vortex limit. In this scenario a single-vortex, mean-field approach is usually profitably exploited to determine the vortex parameters,\cite{WuPRL90}$^-$\cite{PompeoJAP08} such as the viscous drag coefficient, the depinning frequency, the pinning constant. Mean field models have been often extended with the introduction of field-dependent effective vortex parameters in order to account for inhomogeneities and finite elasticity of the vortex system.\cite{OwliaeiPRL92,YehPRB93,WuPRL95,PowellCJP96,GhoshSUST97} 
Mean field models can also be generalized rather easily to include the screening effects and the interactions between the microwave current and moving vortices.\cite{CC,MStheory,Dulcic93} 
As a matter of fact, most experiments concerning the mixed-state electrodynamic properties in the microwave range make use of mean-field models for the interpretation of the data and for the determination of the physics of the vortex matter.

Due to the complexity of microwave measurements, high sensitivity is usually obtained by using resonating techniques,\cite{PetersanJAP98} while wideband measurements (e.g. Corbino disk,\cite{Corbino,BoothRSI65,WuCorbino,CorbinoG20} bolometry \cite{bolo}) are confined to small temperature regions. Thus, one has to deal with single-frequency measurements. Discussion of such measurements in terms of vortex parameters is only seemingly straightforward. In fact, it is quite delicate because the choice of the specific model can deeply influence the figures which can be extracted, since a different dynamics (i.e., frequency dependence) of the data with respect to the model adopted can significantly vary the estimates of the vortex parameters (a discussion, in a different context, is given in Refs.\onlinecite{ParksPRL95,ParksJPCS95}).

An example is given in Ref.\onlinecite{Janjusevic06}, which presents measurements on Nb films: the flux flow resistivity as calculated from the data displays very different field dependencies whether the pinning contribution is taken into account or not.
Another example, specific to HTCS, is discussed in the review of Ref.\onlinecite{GolosovskySUST96}: there, it is shown that estimates of the viscous drag coefficient can vary up to two orders of magnitudes whether pinning effects are included or not. In addition to those remarks, one has to note that in HTCS thermal creep is unavoidable, at least above $\sim T_c/2$: without a proper data treatment, the use of even the simplest vortex models would pose serious questions on the reliability of the estimates of the vortex parameters.

From the above discussion, it is clear that the reliability of the determination of the vortex parameters from experimental data, and an evaluation of their model-dependence, is an important issue before those parameters can be compared to microscopic theories. The fact that in most models the number of parameters exceeds the number of independently measured observables further complicates the problem.

In this paper we intend to address the issue of a correct and reliable determination of the vortex parameters based on typical microwave measurements. In so doing, we will show first that many mean-field models can be reduced to a universal expression for the vortex resistivity. We will use this result to discuss the model-independence of the vortex parameters. Second, by exploiting physical and algebraical properties of the models for the vortex complex resistivity, we will show that many additional information, including creep, can be extracted with respect to standard analysis. Finally, we will apply the novel, extended analysis to some illustrative example.

The outline of this paper is as follows. In Section \ref{sec:models} we briefly recall the main mean field models for the vortex motion, with emphasis on their common features and differences, and we rewrite the resulting expressions for the complex resistivity in a single formulation.
Simple numerical examples are used to illustrate the lack of robustness of the derivation of the vortex parameters with respect to different models. In Section \ref{sec:analysis} selected data of the microwave magneto-resistivity in the mixed state of some HTCS thin films, Tl$_2$Ba$_2$CaCu$_2$O$_{8+x}$ (TBCCO) and YBa$_2$Cu$_3$O$_{7-x}$ (YBCO), are discussed in detail in the light of the previous Section, to exemplify the proposed extended analysis. The results arising from the application of different models are discussed. In Section \ref{sec:summary} we summarize the results and prospect future work. The Appendices report most of the details of the calculations.

\section{The mixed state microwave complex resistivity}
\label{sec:models}
In this Section we consider the complex vortex resistivity $\rho_{vm}$ as calculated within mean field models. We focus on the local response of the fluxon system. In this case, the surface impedance $Z$ can be derived  when needed by properly taking into account the geometry of the sample and of the experiment.\cite{surfImp}
We specifically discuss the common case where the microwave currents flow in the isotropic plane [$(a,b)$ plane, in cuprates] of a uniaxial anisotropic superconductor, and the static magnetic induction field $\mathbf{B}$ is perpendicular to the plane of isotropy and to the alternate currents. 
The vortex motion resistivity $\rho_{vm}$ relates the alternate current density $\mathbf{J}$ to the electric field $\mathbf{E}=\mathbf{B}\times\mathbf{v}$ induced by the moving vortices.

When the vortex displacement is sufficiently small to ensure the validity of the local limit, e.g. with increasing the driving frequency, the vortex velocity $\mathbf{v}$ is customarily determined by writing down the dynamic equation for the balance of forces (per unit length) exerted on a single vortex:\cite{BS, GR, Martinoli, WangSUST96, GolosovskySUST96}
\begin{equation}
\label{eq:forces}
    \eta\mathbf{v}+\alpha_H\mathbf{\hat{n}}\times\mathbf{v}+\nabla U=
    \mathbf{J}\times\mathbf{\hat{n}}\Phi_0+\mathbf{F_{thermal}}
\end{equation}
\noindent where $\mathbf{\hat{n}}$ is the unit vector along the vortex and $\Phi_0$ is the flux quantum.
The overall current density $\mathbf{J}$ exerts the Lorentz force $\mathbf{J}\times\mathbf{\hat{n}}\Phi_0$. Thermal fluctuations give rise to the stochastic force $\mathbf{F_{thermal}}$, responsible for fluxon jumps between pinning sites.

Power dissipation of moving vortices is represented by the drag force $\eta \mathbf{v}$, where the viscous drag coefficient $\eta$ is related to the relaxation processes of the quasiparticles.\cite{kopninRPP} Thus, it is a quantity essentially connected to the microscopic electronic state. The viscous drag coefficient $\eta$, which is also commonly referred to as vortex viscosity, is not to be confused with the vortex fluid viscosity discussed in Ref.\onlinecite{marchettiPRB90} and there denoted with the same symbol $\eta$. The symbol $\eta$ is here chosen following the common use.

The perpendicular (Hall) force on a moving vortex, $\alpha_H\mathbf{\hat{n}}\times\mathbf{v}$, is described by the Hall coefficient $\alpha_H$. The Hall angle is $\theta_H=\arctan(\alpha_H/\eta)$.

The effects of pinning are represented by the force $-\nabla U$ where $U$ is the spatial function describing the pinning potential.

In the harmonic regime ${e^{\rmi\omega t}}$, at sufficiently high angular frequency $\omega$ only very small oscillations, around equilibrium positions, are involved. Thus, one approximates $\nabla U \simeq k_p\mathbf{v}/(\rmi \omega)$, where $k_p$ is the pinning constant (also indicated as the Labusch parameter $\alpha_L$) and $\mathbf{v}/(\rmi \omega)$ is the vortex displacement. This force is the result of single interactions between pins and vortices, of collective interaction of the ensemble of pins and the vortex matter, and of the fluxon system elasticity itself. As such, it is clear that it contains much of the physics of the vortex matter.

In the left-hand side of Eq.\eqref{eq:forces} the balance of the various forces acting on a moving vortex is strongly frequency dependent. An important characteristic frequency is the so-called (de)pinning angular frequency $\omega_p=k_p/\eta$, which marks the crossover between elastic motion, dominant at lower frequencies, from purely dissipative motion, arising at higher frequencies.

The role of vortex mass is a longstanding issue, \cite{massDebate_small,massDebate_large} that seems far from an accepted solution. Accordingly to most estimates, \cite{massExtimations_negligible} in the following we will neglect the vortex mass for the microwave frequency range we are interested in.

Equation \eqref{eq:forces} is the starting point for many models. Different expressions of $\rho_{vm}$ have been derived depending on the specific physics incorporated in the various terms of Eq.\eqref{eq:forces}. As important examples, the inclusion of finite vortex elasticity and/or collective pinning phenomena through properly defined field dependent pinning parameters\cite{WuCorbino,PowellCJP96,GhoshSUST97,YehPRB93} yielded to a much wider application of Eq.\eqref{eq:forces}.

In the following, we briefly recall several specific models for the vortex resistivity, and we show that all can be reduced to a single analytic expression.\\

\textit{Gittleman and Rosenblum (GR) model.}~---
In this seminal work\cite{GR} no thermal and Hall terms were considered: $\mathbf{F_{thermal}}=$0, $\alpha_H=$0 in Eq.\eqref{eq:forces}. Thus:
\begin{equation}
\label{eq:rhoGR}
    \rho_{vm,GR}=\frac{\Phi_{0}B}{\eta}\frac{1}{1-\rmi\frac{\omega_{p}}{\omega}}
\end{equation}
\noindent In this model $\eta$ and $\omega_p$ can be directly calculated from the data by simple inversion. In the high frequency limit ($\omega\gg\omega_p$) $\rho_{vm,GR}\rightarrow\rho_{ff}$, being $\rho_{ff}=\Phi_{0}B/\eta$ the free flux flow resistivity. Equation \eqref{eq:rhoGR} gave for many years the theoretical grounds for the interpretation of microwave and radiofrequency (rf) data, and it has served as an essential interpretative tool for many experiments performed in HTCS. \cite{GRciteLTS,GRciteHTS,GhoshSUST97,WuPRL90, HebardPRB89,HuangPhC92, MorganPhC94, GolosovskyPRB94}

We note here the relevance of the so-called $r$-parameter: $r=\frac{\Im(\rho_{vm,GR})}{\Re(\rho_{vm,GR})}$. In general, it gives a measure of the relative weight of the reactive to resistive response, given essentially by the elastic and dissipative response of the vortex. However, in this specific model it directly yields $r=\omega_p/\omega$. The peculiar, quantitative role that $r$ plays in the GR model, together with the fact of being an experimental quantity, make it an important parameter in the discussion of the data (see also the thorough discussion in Ref.\onlinecite{HalbritterJap90}).\\

\textit{Coffey-Clem (CC) model.}~---
By considering nonzero $\mathbf{F_{thermal}}$, and assuming a sinusoidal pinning potential $U(x)$, Coffey and Clem calculated the vortex resistivity as:\cite{CC}
\begin{equation}
\label{eq:rhoCC}
    \rho_{vm,CC}=\frac{\Phi_{0}B}{\eta}\frac{\epsilon+\rmi\frac{\omega}{\omega_{0}}}{1+\rmi\frac{\omega}{\omega_{0}}}
\end{equation}
\noindent where $\epsilon$ is a dimensionless creep factor. In the assumption of a uniform periodic pinning potential of height $U_0$ and by defining a normalized energy barrier height $u=U_0(T,B)/(K_B T)$ ($K_B$ is the Boltzmann constant), the following expressions hold:
\begin{equation}
\label{eq:epsilon}
    \epsilon=[I_0(u/2)]^{-2}
\end{equation}
\begin{equation}
\label{eq:nu0}
    \omega_0=\omega_p\frac{1}{1-\epsilon}\frac{I_1(u/2)}{I_0(u/2)}
\end{equation}
\noindent where $I_0$ and $I_1$ are the modified Bessel functions of the first kind and orders 0 and 1, respectively. According to
Eq.\eqref{eq:epsilon}, $0\leq\epsilon\leq1$.  In the limit $\epsilon\rightarrow 0$, $\omega_{0}\rightarrow\omega_{p}$ and Eq.\eqref{eq:rhoCC} reverts to the simpler GR model. For $\epsilon\rightarrow 1$ ($U_0\rightarrow 0$), as it can happen near the critical temperature $T_c$, thermal creep completely washes out pinning so that $\rho_{vm}$ reverts to pure flux flow $\rho_{ff}=\Phi_0 B/\eta$ at all frequencies.\\

\textit{Brandt (B) model.}~---
Brandt considered the creep effects by introducing a phenomenological thermally relaxing pinning constant $k_{p,t}(t)=k_{p}e^{-t/\tau_r}$ (Ref.\onlinecite{brandtkp}) so that the corresponding vortex resistivity is:\cite{brandt}
\begin{equation}
\label{eq:rhoB}
    \rho_{vm,B}=\frac{\Phi_{0}B}{\eta}\frac{\epsilon'+\rmi\omega\bar{\tau}}{1+\rmi\omega\bar{\tau}}
\end{equation}
\noindent where $\epsilon'=\frac{\tau_p}{\tau_p+\tau_{r}}$ takes the role of a creep parameter,
$\bar{\tau}=\frac{\tau_p\tau_{r}}{\tau_p+\tau_{r}}$, $\tau_{r}=\tau_p e^{U_0/K_BT}$, being $\tau_p=1/\omega_p$ the usual (de)pinning characteristic time and $U_0$ the pinning potential barrier height.
It is worth noting that, within this model, a tighter algebraical condition $0\leq\epsilon'\leq1/2$ holds. This model does not require any specific assumption about the pinning potential landscape, thus allowing in principle to include also other mechanisms like quantum flux creep.\cite{brandtone} On the other hand, it cannot be used for too high creep rates ($U_0\rightarrow 0$ $\Leftrightarrow$ $\epsilon'\rightarrow 0.5$), since in this limit it does not recover the expected $\rho_{ff}$ limit for any finite or zero frequency.\\

\textit{Two-mode (TM) model}~---
The GR, CC and B models considered a single oscillatory mode for the vortex motion. A second dynamic mode of the fluxon system arises from the bending of the flux lines.\cite{SoninTheory,MStheory} This additional mode strongly affects $\rho_{vm}$ in the case of surface pinnng, that adds up to the bulk pinning. In the present case of thin films, in absence of thermal activation the following expression takes place:\cite{MSexp}
\begin{equation}
\label{eq:rhoMS}
    \rho_{vm,TM}=\frac{\Phi_{0}B}{\eta}\frac{1}{1-\rmi\frac{\omega_{p,eff,TM}}{\omega}}
\end{equation}
\noindent where $\omega_{p,eff,TM}=k_p/\eta+2\varepsilon_l\Phi_0/(l \eta d)=\omega_p+\omega_{ps}$ is an overall pinning frequency combining the effects of bulk pinning ($\omega_p$) and surface pinning ($\omega_{ps}$), $d$ is the thickness of the thin film, $\varepsilon_l$ is a (field dependent) fluxon line energy,\cite{MStheory} and $l$ is a characteristic length describing the surface pinning.\cite{MSexp} When surface pinning is negligible, $l\rightarrow\infty$. Equation \eqref{eq:rhoMS} coincides with the GR expression, Eq.\eqref{eq:rhoGR}, once the pinning constant is redefined with the additional contribution given by surface pinning. By contrast, for bulk geometries (beyond the scope of this work) one has to take into account the bending of flux lines and the redistribution of the currents, so that the full expression becomes rather complex.\cite{MStheory}\\

\textit{Universal expression.}~---
Despite their different approaches, all the previously recalled models for $\rho_{vm}$ can be cast in the following single analytical expression:
\begin{equation}
\label{eq:rhovm}
    \rho_{vm}=\rho_{vm,1}+\rmi\rho_{vm,2}=\frac{\Phi_{0}B}{\eta_{eff}}\frac{\epsilon_{eff}+\rmi\omega\tau_{eff}}{1+\rmi\omega\tau_{eff}}
\end{equation}
\noindent where $\tau_{eff}$ is the main time constant governing the vortex oscillations (usually linked to the pinning characteristic frequency), and the dimensionless parameter $\epsilon_{eff}$ is a measure of the weight of creep phenomena. It is bound to the interval [0,1] in order to recover the correct zero frequency limit, which must be $0\leq\rho_{vm}\leq\rho_{ff}$. We note that, according to Eq.\eqref{eq:rhovm}, $\rho_{vm,2}\geq0$ always.

\textit{Inclusion of the Hall term.}~---
For completeness, we mention that also the Hall term can be incorporated in the models for $\rho_{vm}$. Neglecting creep for the sake of simplicity, the straightforward extension of the GR model yields:\cite{ParksHall}
\begin{equation}
\label{eq:rhoGRH1}
    \rho_{vm,GRH}=\frac{\Phi_{0}B}{\eta}\frac{1-\rmi\frac{\omega_{p}}{\omega}}{\left(1-\rmi\frac{\omega_{p}}{\omega}\right)^{2}+\tan^2\theta_H}
\end{equation}
\noindent valid at constant microwave currents (it is useful to recall that most of cavity/resonator experiments meet this condition, while e.g. Corbino disk experiments do not).
The Hall term introduces a different frequency dependence: a Bode analysis reveals a pole in zero, as in the previously discussed models, and two additional complex poles which introduce a resonant response. Nevertheless, one can formally cast Eq.\eqref{eq:rhoGRH1} in the form of the universal expression, Eq.\eqref{eq:rhovm}, with $\epsilon_{eff}=0$ and defining an effective viscous drag coefficient $\eta_{eff,H}=\eta+\alpha^2/\eta$ - often considered in microwave experiments \cite{GolosovskySUST96,TsuchiyaPRB01} - and
\begin{equation}
\label{eq:nupH}
    \frac{1}{\tau_{eff}}=\omega_{p,eff,H}=\omega_p\frac{1+\left(\frac{\omega_p}{\omega}\right)^2-\tan^2\theta_H}{\left(1+\left(\frac{\omega_p}{\omega}\right)^2\right)\left(1+\tan^2\theta_H\right)}.
\end{equation}
\noindent $\omega_{p,eff,H}$ formally plays the role of an effective pinning frequency, but retains a dependence from the measuring frequency $\omega$. For not too large Hall angles (i.e., for $\tan\theta_H<\pi/4$, which is a condition generally expected in cuprates \cite{TsuchiyaPRB01,HanaguriPRL99,MaedaPC01}), $\omega_{p,eff,H}$ is always a positive quantity. When interpreting single-frequency measurements, one has to bear in mind that the presence or absence of the Hall term might not be an obvious issue. At very high measuring frequency (in the THz range) one should take the Hall term into account.\cite{ParksPRL95}

The reduction of many models to the single analytical expression, Eq.\eqref{eq:rhovm}, allows to conceptually divide the whole data analysis and interpretation process into two steps.

First, using the sole assumption of standard vortex dynamics, one can derive from the experimental data several model-independent quantities: $\eta_{eff}, \tau_{eff}, \epsilon_{eff}$. Although their physical meaning can be fully determined only with the choice of a particular model, their determination relies only upon Eq.\eqref{eq:rhovm}, and thus it applies indifferently to all the specific models described by Eq.\eqref{eq:rhovm}.

Second, the full physical interpretation of the effective parameters is given after the choice of a specific vortex model. This last step can lead to quite different results, hence its delicate nature.

Accuracy issues, applicability limits, general constraints are thus interesting for the analysis of the experiments.\\

\textit{Numerical examples.}~---
We now illustrate with two numerical examples the effect of neglecting flux creep in the interpretation of the data, and then the need for a more accurate data treatment. To do so, we first generate data for $\rho_{vm}(B)$ using complete models (e.g., the CC or B models) and reasonable (according to the literature) values of parameters. Then, we use a simplified model (GR) to evaluate $\eta$ and $k_p$ from the as-generated data. We show that this procedure, which mimics a very widely used approach to the interpretation of the experiments, can (1) yield wrong estimates for the vortex parameters and (2) drive toward a complete misinterpretation of the physics at the origin of experimental data.

In the first example we compute $\rho_{vm}(B)=\rho_{vm,1}(B)+\rmi \rho_{vm,2}(B)$ at 25 GHz by means of the B model, selecting parameters appropriate for a large elastic response (as, e.g., in YBCO with strong pinning below 80 K): $\eta=10^{-7}$ Ns/m$^2$, $k_p=3\times10^4$ N/m$^2$ and a small but finite creep factor, $\epsilon'=0.15$.
\begin{figure}[htb]
  % Requires \usepackage{graphicx}
\centerline{\includegraphics[width=7.5cm]{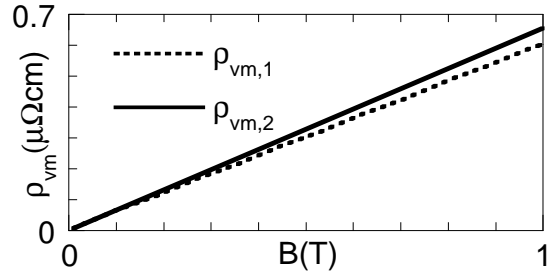}}
  \caption{Simulated $\rho_{vm}(B)$ vs field through the B model ($\eta=10^{-7}$ Ns/m$^2$, $k_p=3\times 10^4$ N/m$^2$, $\epsilon'=0.15$)  representing an YBCO with strong pinning, $T\sim$80 K and $\nu$=25 GHz.}
\label{fig:sim1}
\end{figure}

As seen in Fig.\ref{fig:sim1}, both $\rho_{vm,1}$ and $\rho_{vm,2}$ are linearly proportional to the field $B$, which is usually taken as an indication for the absence of creep effects. However, applying the GR model to these same $\rho_{vm}(B)$ data one obtains $\eta= 1.57\times10^{-7}$ Ns/m$^2$, largely different from the ``true" value. By contrast, one would have $k_p=2.7\times10^4$ N/m$^2$, 10\% far from the true value. It is noteworthy that neglecting creep has the largest influence on the estimate of $\eta$. This point will be further investigated later.

\begin{figure}[thb]
  % Requires \usepackage{graphicx}
\centerline{\includegraphics[width=7.5cm]{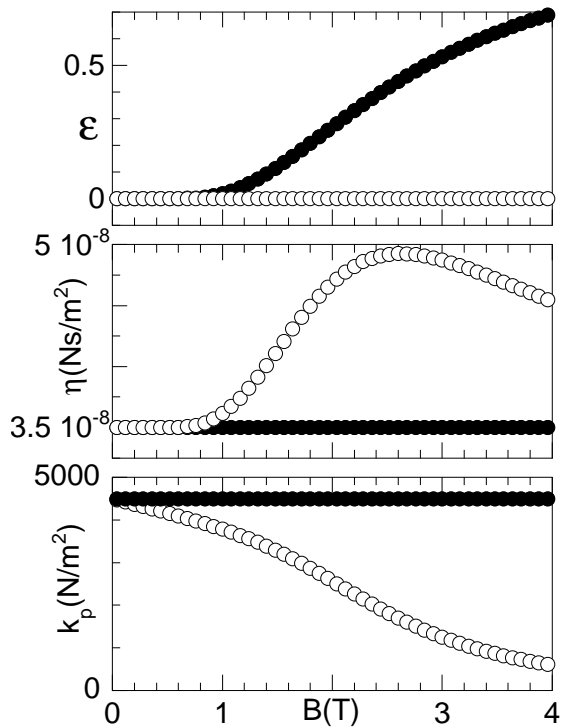}}
  \caption{Vortex parameters used to reproduce real data\cite{WuCorbino} of an YBCO thin film at 83.5 K and $\nu$=13.03 GHz. Full symbols: values proposed in Ref.\onlinecite{WuCorbino} through CC model fits ($k_p=4.5\times 10^3$ N/m$^2$, $\eta=3.5 \times 10^{-8}$ Ns/m$^2$, $u(B)/2=3.5$ T$/B$). Open circles: values calculated through GR model.}
\label{fig:sim2}
\end{figure}

In the second example we compute $\rho_{vm}(B)$ at 13.03 GHz and 83.5 K by means of the CC model, using parameters that were found to describe excellently real data taken on a YBCO thin film:\cite{WuCorbino} $k_p=4.5\times10^3$ N/m$^2$, $\eta=3.5\times10^{-8}$ Ns/m$^2$, and a field-dependent creep parameter (upper panel of Fig.\ref{fig:sim2}) calculated with a normalized pinning barrier $u(B)/2=3.5$ T$/B$.
Applying the GR model (which neglects flux creep) to the as-generated data, one would find fictitious field dependences in $\eta$ and $k_p$ (center and lower panels of Figure \ref{fig:sim2}, open circles). We note that the fictitious field dependence of $k_p(B)$ is particularly dangerous for a proper interpretation: although completely artificial, it is potentially verisimilar, since it could be referred to a change in the pinning of the fluxon system.\cite{GR,LO}
Thus, the choice of the model has a very strong impact on the evaluation and interpretation of the vortex parameters. In the following Section we illustrate a method for the analysis of the data which alleviates this issue, giving reliable and possibly model-independent estimates of the vortex parameters and, most important, estimates of the error bars. To keep contact with experiments, we illustrate the procedure making use of measurements of the complex magneto-resistivity on various HTCS samples.

\section{Extended analysis of experimental data}
\label{sec:analysis}

\subsection{Experimental data}
\label{sec:experimental}
\begin{figure}[h]
  % Requires \usepackage{graphicx}
\centerline{\includegraphics[width=7.5cm]{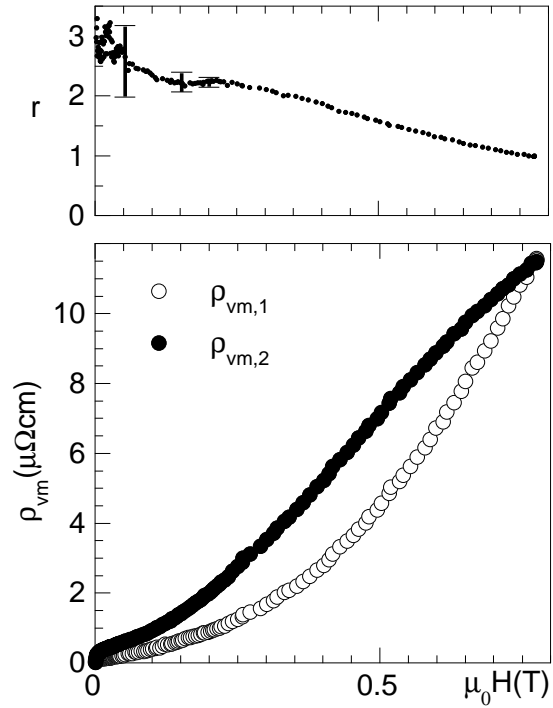}}
  \caption{Vortex resistivity $\rho_{vm}$ vs applied field $H$ in the TBCCO sample at $T$ = 81.4 K ($T/T_c\simeq$ 0.782). Upper panel:  parameter  $r=\rho_{vm,2}/\rho_{vm,1}$. Lower panel: $\rho_{vm,1}$ and $\rho_{vm,2}$.}
\label{fig:DeltaZT}
\end{figure}

\begin{figure}[h]
  % Requires \usepackage{graphicx}
\centerline{\includegraphics[width=7.5cm]{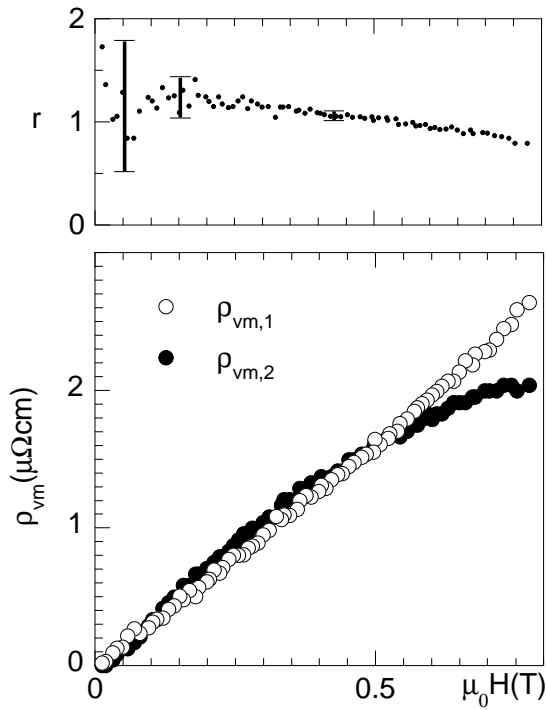}}
  \caption{Vortex resistivity $\rho_{vm}$ vs applied field $H$ in the YBCO sample at $T$ = 70.9 K ($T/T_c\simeq$ 0.787). Upper panel:  parameter  $r=\rho_{vm,2}/\rho_{vm,1}$. Lower panel: $\rho_{vm,1}$ and $\rho_{vm,2}$.}
\label{fig:DeltaZY}
\end{figure}

The measurements here used for illustration were taken on a TBCCO thin film ($T_c\simeq104$ K) and on a YBCO thin film ($T_c\simeq90$ K) with BaZrO$_3$ inclusions. Details of the film preparation have been given elsewhere.\cite{tallioPrep,GalluzziIEEE07}
Microwave measurements were performed by means of a sapphire dielectric resonator\cite{dielRes} at $\nu=47.7$ GHz, with a static magnetic field $\mu_0 H\simeq B$ perpendicular to the surface of the samples (aligned with the $c$ axis). The variation of the microwave resistivity with the field yielded the vortex motion contribution, $\rho_{vm}$.
Typical measurements of $\rho_{vm}(H)$ at similar reduced temperatures for both samples are shown in Figures \ref{fig:DeltaZT} and \ref{fig:DeltaZY}.
The quantity $r=\rho_{vm,2}/\rho_{vm,1}$ is also shown, given its relevance in the analysis. Due to unavoidable numerical uncertainties in the calculation of $r$ at low fields, we will limit the analysis of the data at fields  $\mu_0 H>0.1$ T.

In both samples a significant magnitude of the reactive component can be observed: $r\gtrsim1$. In TBCCO, the upward curvature in the data (more evident in $\rho_{vm,1}$) suggests field dependent vortex parameters, with a possibly relevant flux-creep. In YBCO $\rho_{vm,1}(H)$ is almost perfectly linear in $H$, consistent with a simpler scenario of constant viscous drag coefficient and negligible flux-creep. We note that $\rho_{vm}$ in TBCCO is much larger than in YBCO at the same fields, by about one order of magnitude.
\subsection{Conventional analysis}
\label{sec:conventional}
\begin{figure}[h]
  % Requires \usepackage{graphicx}
\centerline{\includegraphics[width=7.5cm]{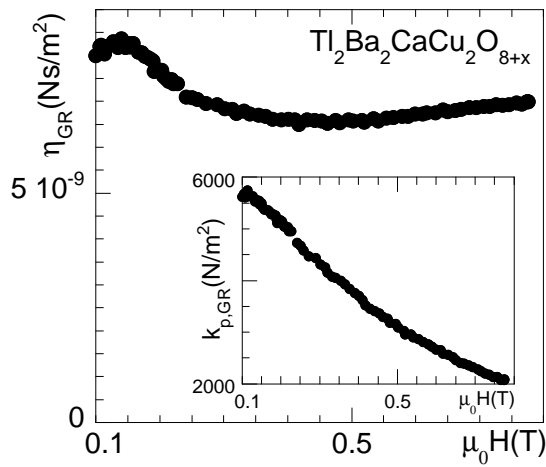}}
  \caption{Vortex parameters as derived within the GR model in TBCCO at 81.4 K. Main panel: viscous drag coefficient $\eta_{GR}$. Inset: pinning constant $k_{p,GR}$.}
\label{fig:ParsNoCreepT}
\end{figure}

\begin{figure}[h]
  % Requires \usepackage{graphicx}
\centerline{\includegraphics[width=7.5cm]{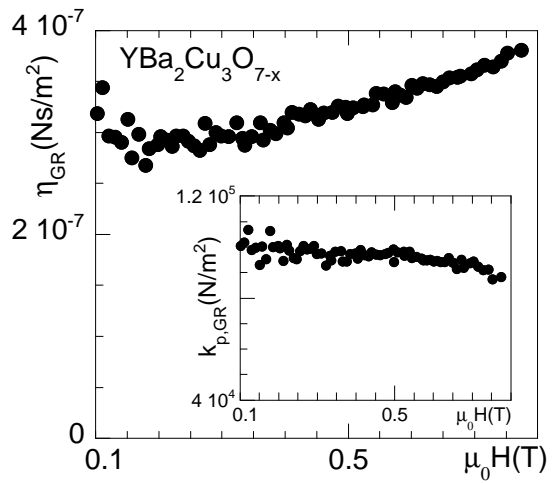}}
  \caption{Vortex parameters as derived within the GR model in YBCO at 70.9 K. Main panel: viscous drag coefficient $\eta_{GR}$. Inset: pinning constant $k_{p,GR}$.}
\label{fig:ParsNoCreepY}
\end{figure}

We first apply the conventional approach (GR model, no flux-creep) to our data. By inverting Eq.\eqref{eq:rhoGR}, $\eta_{GR}$ and $k_{p, GR}$ are directly obtained from the data (the subscript ``GR'' stands for zero-creep derived values), and are reported in Figures \ref{fig:ParsNoCreepT} and \ref{fig:ParsNoCreepY} for TBCCO and YBCO, respectively.
In YBCO $k_{p,GR}$ is constant and $\eta_{GR}$ shows a weak field dependence (approximately within 15\% of the average value). Absolute values are in agreement with the literature (see Ref.\onlinecite{GolosovskySUST96} and references therein). Apart from minor considerations, these results seem in line with conventional vortex dynamics, and would not stimulate particular comments.
By contrast, in TBCCO $\eta_{GR}$ is approximately field-independent but attains very small values, about one order of magnitude smaller than typical values in YBCO. In addition, $k_{p,GR}$ decreases markedly (by a factor $\sim$3) with the field. However, as previously shown, those features could well be an artifact of the simplified analysis. Since both features might represent non-standard vortex dynamics, we now illustrate a method to assess the reliability of the calculations of the vortex parameters and, as a consequence, of the physics that can be derived from them.
\subsection{Extended analysis}
\label{sec:extended}
We first exploit the analytical and physical features of the universal expression, Eq.\eqref{eq:rhovm}. Only as a second step we make use of specific models.\\

{\it Creep factor}~--- The creep factor $\epsilon_{eff}$ cannot be directly obtained from the data. However, it is possible to give an upper limit $\epsilon_{eff,max}$ according to the following expression,\cite{tallioJS} derived in Appendix A (we recall that $r$ is an experimental quantity): 
\begin{equation}
\label{eq:defChimax1}
    \epsilon_{eff}\leq\epsilon_{eff,max}(r)=1+2r^2-2r\sqrt{1+r^2}
\end{equation}
This is an important result, that will be used throughout the remaining of this paper.

\begin{figure}[h]
  % Requires \usepackage{graphicx}
\centerline{\includegraphics[width=7.5cm]{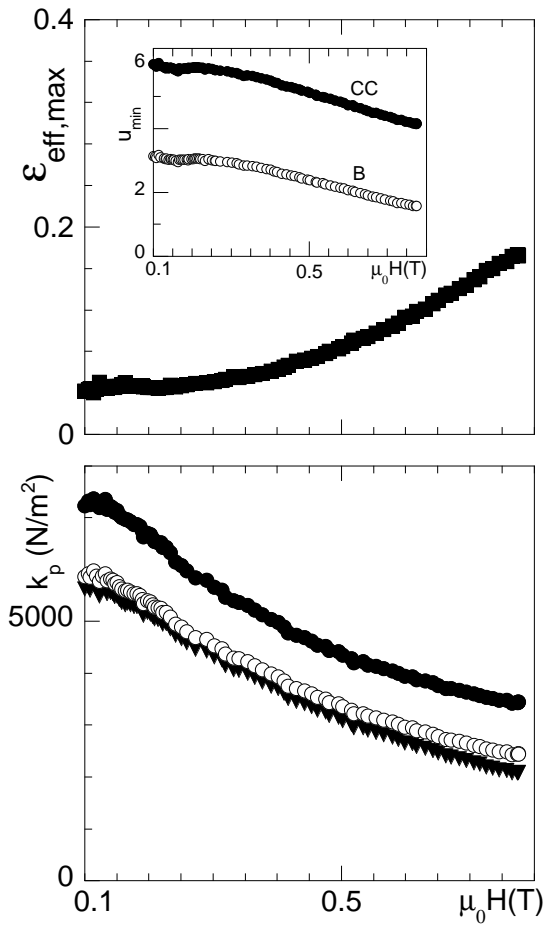}}
  \caption{Model dependence of vortex parameters in TBCCO at 81.4 K. Upper panel: maximum creep factor $\epsilon_{eff,max}(H)$. Inset: corresponding minimum barrier energies $u_{min}$ of the pinning potential within the CC (full dots) and B (open symbols) models. Lower panel: $k_p$ within the CC (maximum values, full dots), B (maximum values, open symbols) and GR (down-triangles) models.}
\label{fig:chiMaxT}
\end{figure}

As an example, $\epsilon_{eff,max}(B)$ at $T$=81.4 K for the TBCCO sample is plotted in Fig.\ref{fig:chiMaxT} (upper panel). It can be seen that $\epsilon_{eff,max}$ is an increasing function of $B$. In addition, we have found that it is also an increasing function of $T$, consistently with the thermal origin of this parameter.\\

{\it Viscous drag coefficient}~--- An important constraint on the viscous drag coefficient $\eta$ -which we recall is also referred to as vortex viscosity- can be derived (calculations are reported in Appendix A): $\eta$ always falls in the range
\begin{equation}
 \eta_{GR}\frac{1+\epsilon_{eff,max}}{2} < \eta < \eta_{GR}
\label{eq:rangeta}
\end{equation}
Note that Eq.\eqref{eq:rangeta} gives model-independent (in the meaning of ``all the dynamics that can be described by Eq.\eqref{eq:rhovm}") ranges in terms of experimental quantities: $\eta_{GR}$ and $\epsilon_{eff,max}$ are directly calculated from experimental data.

The values obtained through the use of the GR model always represent an upper limit for $\eta$. Eq.\eqref{eq:rangeta} gives a counterintuitive result: the GR estimate, $\eta_{GR}$, turns out to be a more precise approximation to $\eta$ when $\epsilon_{eff,max}$ is large, most likely at high temperature. In other words, {\it in presence of a large reactive component -i.e. large $r$- the application of the GR model to low temperature data, where creep is thought to be small or neglibible, yields more uncertain estimates of $\eta$}. In this case it is true that $\epsilon_{eff,max}$ is small (and that the real $\epsilon_{eff}$ can be even smaller), but neglecting it altogether brings an error up to a factor of 2 on $\eta$. It is not excluded that the present result might bring a reconsideration of many of the values obtained for $\eta$ in the literature.

Thus, estimates for $\eta$ can be accompanied by significant error bars. Errors can be reduced by noting that the uncertainty on the creep factor,  $0 < \epsilon_{eff} < \epsilon_{eff,max}$, does not uniformly reflect on $\eta$. 
In Appendix B we discuss in detail a statistical approach. The main result, that we state here, is that even in the worst case $\langle\eta\rangle\pm20\%$ represents a 90\% confidence interval, where $\langle\eta\rangle$ is the expected value of $\eta$.

We stress again that the results derived up to now stem from the general expression,  Eq.\eqref{eq:rhovm}, and thus are independent from the specific model adopted. In particular, the experimental data here reported in TBCCO actually indicate a very small viscous drag coefficient.\cite{notaHall}\\

{\it Activation energy.}~--- From the maximum creep factor $\epsilon_{eff,max}$ it is possible, in principle, to derive a \emph{minimum} barrier energy $U_{0,min}$ of the pinning potential. To do so, a specific model must be chosen. The widely used CC and Brandt models usually yield similar results, in the range where the Brandt model can be applied ($\epsilon_{eff,max}<0.5$, that is $r\gtrsim0.35$), as reported in the inset of the upper panel of Fig.\ref{fig:chiMaxT}. The main difference is a roughly vertical translation, arising from the different treatment of the thermal creep contribution.

As an illustration, in TBCCO we obtain $U_{0,min}(B$=0.2 T$)\simeq$20 meV and 40 meV in the CC and Brandt model, respectively. Similar values have been reported in YBCO, as determined through microwave multifrequency measurements.\cite{P&P,Belk}\\

{\it Pinning constant.}~--- The uncertainty in the determination of $k_p$ behaves in the opposite way with respect to the viscous drag coefficient $\eta$: the GR value $k_{p,GR}$ represents the lower bound, whereas finite values of $\epsilon_{eff}$ yield larger $k_p$, up to a maximum value $k_{p,max}$ which is model dependent. An interpretation of the data through the Brandt model yields $k_{p,max}\simeq k_{p,GR}$, thus giving very small uncertainties (see comments in Appendix B). However, an interpretation of the same data within the CC model yields a larger uncertainty, as represented in lower panel of Fig.\ref{fig:chiMaxT}. Care must be taken before driving conclusions from an analysis of $k_p$: the model chosen -differently from $\eta$- plays a very important role in the derivation of $k_p$ from the data.
Only the lower bound is model independent.
The uncertainty in $k_p$ has a trend opposite with respect to $\eta$. As exemplified in Fig.\ref{fig:chiMaxT}, the relative uncertainty grows as $k_p$ decreases, which coincides with decreasing $r$. 

Coming back to the illustrative data reported here in TBCCO we observe that, whatever the model chosen (CC or B), the uncertainty band of allowed values is narrow enough to unambigously determine that $k_p$ is a decreasing function of the field.
The study of the physics involved in this field dependence is beyond the scope of this work, and it will be investigated in the future. We only mention that such a behavior may originate from rather different physics. For instance, a finite, small density of effective pinning centers can yield a decreasing average pinning strength over isolated vortices/vortex-bundles as the increasing number of vortices with the field progressively exceeds the number of pins.\cite{GR} Another possible scenario could be related to some field-driven transformation of the fluxon lattice, impacting vortex elasticity properties and, ultimately, pinning.\cite{blatterone}\\

\begin{figure}[h]
  % Requires \usepackage{graphicx}
\centerline{\includegraphics[width=7.5cm]{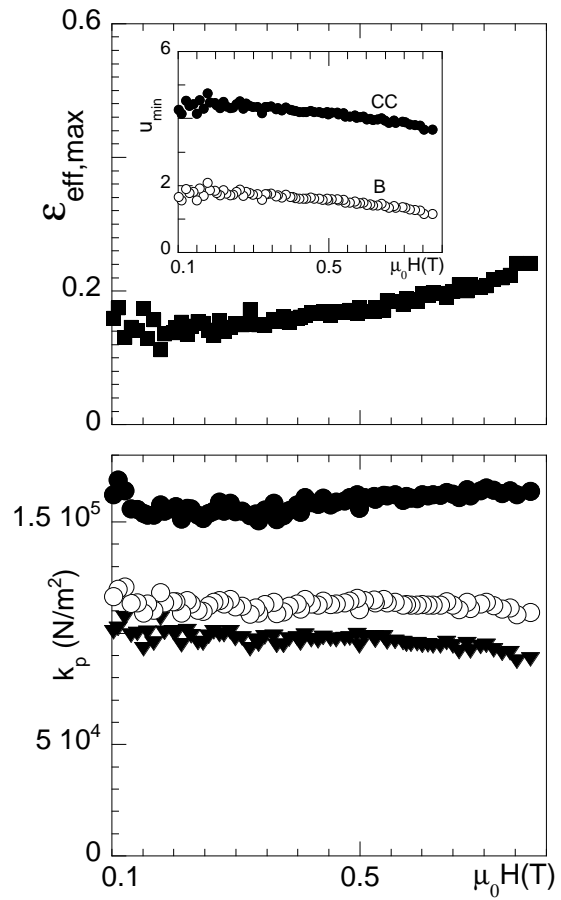}}
  \caption{Model dependence of vortex parameters in YBCO at 70.9 K. Upper panel: maximum creep factor $\epsilon_{eff,max}(H)$. Inset: corresponding minimum barrier energies $u_{min}$ of the pinning potential within the CC (full dots) and B (open symbols) models. Lower panel: $k_p$ within the CC (maximum values, full dots), B (maximum values, open symbols) and GR (down-triangles) models.}
\label{fig:chiMaxY}
\end{figure}

{\it Reanalysis of YBCO data}~--- For completeness, we now consider the data\cite{PompeoAPL07} taken on the YBCO sample reported in Fig.\ref{fig:DeltaZY} and analysed according to the conventional analysis as reported in Fig.\ref{fig:ParsNoCreepY}. 
The extended analysis confirms the results for the viscous drag coefficient: its behaviour is fully compatible with a field independent value, and numerical values are in agreement with literature.
The pinning-related parameters extracted through the extended analysis and reported in Fig.\ref{fig:chiMaxY} add some piece of information. The creep factor $\epsilon_{eff,max}$ and the normalized minimum barrier energy $u_{min}$ exhibit a similar, but weaker, field dependence as in TBCCO. Irrespectively of the choice of the CC or B model,  $k_p$ is compatible with a field-independent behavior, in clear contrast with TBCCO. This fact suggests (a) the existence of pinning wells steeper in YBCO than in TBCCO and (b) the presence of  additional field-induced effects involving the fluxon system in TBCCO.

\section{Summary}
\label{sec:summary}
We have reconsidered the problem of the determination of the vortex parameters from measurements of the complex resistivity. We have reviewed several widely used mean-field models for the vortex dynamics, and we have shown that they can be cast in a single, general expression. We have shown that, independently on the model, neglecting creep effects gives rise to a large uncertainty on the viscous drag coefficient. This property holds true in particular when creep is small (but nonzero). This finding might stimulate a reconsideration of the values for the viscous drag coefficient reported in literature. We have proposed an extended analysis which allows to evaluate and keep under control the uncertainty inherent in the evaluation of the vortex parameters. We have also exploited some model-dependent implications. Overall, with respect to the conventional approaches the analysis here presented allows for the extraction of more, and more controllable, information from the data for the complex resistivity in the vortex state.

\begin{acknowledgments}
We thank H. Schneidewind at IPHT-Jena and G. Celentano at ENEA-Frascati for providing us the TBCCO and YBCO sample, respectively.
\end{acknowledgments}

\section{Appendix A: constraints on vortex parameters.}
\label{sec:algebra}

In this Appendix we consider the universal Eq.\eqref{eq:rhovm} for the vortex resistivity $\rho_{vm}$: using only very general analytical and physical conditions, we derive relations and constraints among the physical quantities involved in the model.

As a first step, we consider the following ratio of experimental quantities in the light of Eq.\eqref{eq:rhovm}:
\begin{equation}
\label{eq:rdef}
    r=\frac{\rho_{vm,2}}{\rho_{vm,1}}=\varpi\frac{1-\epsilon_{eff}}{1+\varpi^{2}\epsilon_{eff}}
\end{equation}
\noindent where $\varpi=1/(\omega\tau_{eff})$. The ratio $r$ belongs to the interval $[0,\infty]$. By rearranging the above relation, one finds:
\begin{equation}
\label{eq:varpi2}
    (r\epsilon_{eff})\varpi^2+(\epsilon_{eff}-1)\varpi+r=0
\end{equation}
\noindent The physical meaning requires $\varpi\geq0$ and $1\geq\epsilon_{eff}\geq 0$ (see Section \ref{sec:models}). Thus, from Eq.\eqref{eq:varpi2} one readily obtains the constraint:
\begin{equation}
\label{eq:defChimax}
    \epsilon_{eff}\leq\epsilon_{eff,max}(r)=1+2r^2-2r\sqrt{1+r^2}
\end{equation}
\noindent which gives an upper bound for the creep factor that can be directly calculated from the data. The function $\epsilon_{eff,max}(r)$ decreases monotonously from 1 to 0 when $r$ spans the interval $[0,\infty]$. The vortex parameters $\eta$ and $\varpi$ are correspondingly delimited within allowed ranges that can be easily determined through straightforward algebra using the explicit expressions for the various vortex parameters reported in Table \ref{tab:limits}.

\begin{table}%[tbp]
\begin{tabular} {ccc}
\hline
&&\\
$\varpi$& = & $\frac{1-\epsilon_{eff}-\sqrt{A}}{2r\epsilon_{eff}}$\\
&&\\
$\eta$& = &$ \frac{\Phi_0 B}{\rho_{vm1}}\frac{2\epsilon_{eff}}{1+\epsilon_{eff}-\sqrt{A}}$\\
&&\\
$\frac{\omega_p}{\omega}$& = &
$\frac{(1-\epsilon_{eff}-\sqrt{A})(1-\epsilon_{eff})}{2r\epsilon_{eff}} \times 
\left\{
\begin{array}{rl}
 \frac{I_0(u/2)}{I_1(u/2)}, & \mbox{CC} \\
\mbox{1}, & \mbox{B}
\end{array}
\right.$\\
&&\\
$u$ & = $\frac{U_0}{K_BT}$ = & $\left\{
\begin{array}{rl}
 2I_0^{-1}(\frac{1}{\sqrt{\epsilon_{eff}}}), & \mbox{CC} \\
&\\
\ln\frac{1-\epsilon_{eff}}{\epsilon_{eff}}, & \mbox{B}
\end{array}
\right.$\\
&&\\
\hline
\end{tabular}
\caption{Explicit expressions for the vortex parameters according to Equation $\eqref{eq:rhovm}$ as a function of $\epsilon_{eff}$ and of the experimental quantities, $\rho_{vm,1}$ and $r$. For the sake of compactness, we defined $A=(1-\epsilon_{eff})^2-4r^2\epsilon_{eff}$. The pinning constant is easily calculated from the definition $k_p=\omega_p/\eta$. 
Here, $I_0^{-1}$ stands for the inverse function of $I_0$.}
\label{tab:limits}
\end{table}

\section{Appendix B: statistical analysis}
\label{sec:statistical}

\begin{figure}[ht]
  % Requires \usepackage{graphicx}
\centerline{\includegraphics[width=7.5cm]{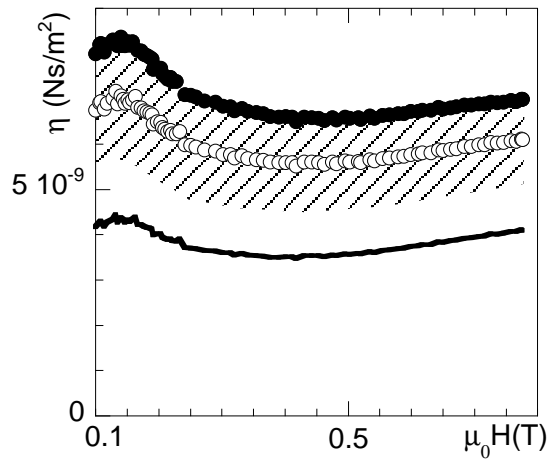}}
  \caption{Estimates of $\eta$ as obtained from the same set of data in TBCCO at $T$=81.4 K. Full symbols: $\eta_{GR}$, assumption of zero creep. Open circles: expected value $\langle\eta\rangle$, assuming that the creep factor has flat probability in its allowed range. Continuous line: minimum allowed value of $\eta$, in the assumption that the creep factor attains the maximum value compatible with the data. Shaded area: 90\% fidelity band, including the zero-creep value.}
   \label{fig:fasciaEta}
\end{figure}

In Appendix A we derived the maximum variability ranges for various vortex parameters. However, more information can be gained by a statistical approach.
We ascribe the maximum uncertainty to $\epsilon_{eff}$: for each ($B$,$T$) point $\epsilon_{eff}$ can be treated as a random variable with rectangular probability density in the $[0,\epsilon_{eff,max}(B,T)]$ range. Thus, other quantities can be described in terms of derived probability densities, computable through standard theorems for the functions of random variables.\cite{papoulis} The expectation value for a generic quantity $a$ derived from $\epsilon_{eff}$ is $\langle a\rangle=\frac{1}{\epsilon_{eff,max}}\int^{\epsilon_{eff,max}}_{0}a(\epsilon_{eff})d\epsilon_{eff}$. The illustrative example of the viscous drag coefficient $\eta$ in TBCCO at $T$=81.4 K (Fig.\ref{fig:fasciaEta}) shows that the expectation value $\langle\eta\rangle$ is closer to the zero-creep value, $\eta_{GR}$, than to the middle of the allowed interval, indicating that large $\epsilon_{eff}$ weakly contribute to $\langle\eta\rangle$. In fact, plotting the $90\%$ fidelity band (shaded area in Fig.\ref{fig:fasciaEta}) starting from the zero creep value $\eta_{GR}$, one notices that the uncertainty on $\epsilon_{eff}$ affects very unevenly the uncertainty of $\eta$. Interestingly, this result relies only on the general expression, Eq.\eqref{eq:rhovm}.

\begin{figure}[ht]
  % Requires \usepackage{graphicx}
\centerline{\includegraphics[width=7.5cm]{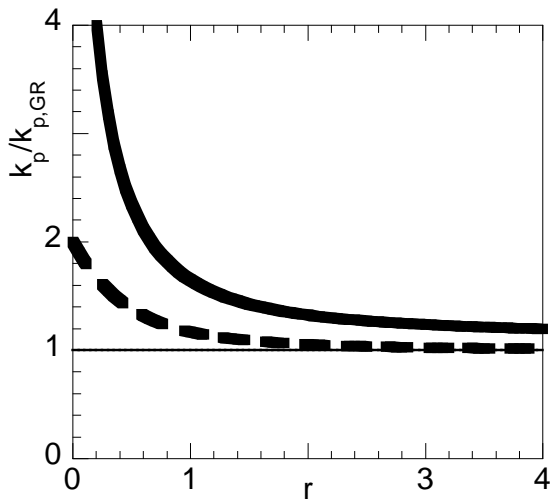}}
  \caption{Dependence of the maximum attainable value of the pinning constant $k_p$ on the experimental $r$ parameter. Calculations are normalized to the GR value, $k_{p,GR}$. A model dependence is evident. Thick line: CC model; dashed line: B model; thin line represent the coincidence with the GR model.}
   \label{fig:fasceKp}
\end{figure}

Similar considerations can be done for the pinning constant $k_p$, but in this case a specific model has to be chosen. We discuss some relevant cases.  Within the Brandt model the uncertainty on $k_p=\omega_p/\eta$ is vanishingly small. Numerically, this is a consequence of the almost exact cancellation of the uncertainties on $\eta$ and $\omega_p$ when $r$ is sufficiently high. This fortunate combination does not take place in the CC model. To illustrate further this effect, Fig.\ref{fig:fasceKp} reports the maximum deviation $k_{p,max}/k_{p,GR}$ as a function of $r$ for both models. It is apparent that for $r\gtrsim1$ the Brandt and GR values are almost coincident ($k_{p,max}/k_{p,GR}\simeq 1$), while the uncertainty for the CC value is larger. It is worth recalling that this favorable property of the Brandt model is counterbalanced by its limitation to regimes with not too high creep rates, whereas the CC approach is able to capture both the low- and the high-creep regimes. 
%

%\section*{References}


\begin{thebibliography}{99}
\bibitem{notaHall} Incidentally, we note that an interpretation of the data by including the Hall term (Eq.\eqref{eq:rhoGRH1}) would lead to an effective viscous drag coefficient $\eta_{eff,H}$ which is again an upper limit for the viscous drag coefficient $\eta$.
\end{thebibliography}
\end{document}